# Range-Max Enhanced Ultra-Wideband Micro-Doppler Signatures of Behind Wall Indoor Human Activities

Qiang An*, Shuoguang Wang*, Ahmad Hoorfar, *Senior Member, IEEE*, Wenji Zhang, Hao Lv, Shiyong Li†,

*Member, IEEE,* and Jianqi Wang†

*Abstract*— Penetrating detection and recognition of behind wall indoor human activities has drawn great attentions from social security and emergency service department in recent years since intelligent surveillance aforehand could avail the proper decision making before operations being carried out. However, due to the influence of the wall effects, the obtained micro-Doppler signatures would be severely degenerated by strong near zero-frequency DC components, which would inevitably smear the detailed characteristic features of different behind wall motions in spectrograms and further hinder the motion recognition and classification. In this paper, an ultra-wideband (UWB) radar system is first employed to probe through the opaque wall media to detect the behind wall motions, which often span a certain number of range bins. By employing such a system, a high-resolution range map can be obtained, in which the embedded rich range information is expected to be fully exploited to improve the subsequent recognition and classification performance. Secondly, a high-pass filter is applied to remove the wall's effects in the raw range map. Then, with the aim of enhancing the characteristic features of different behind wall motions in spectrograms, a novel range-max enhancement strategy is proposed to extract the most significant micro-Doppler feature of each time-frequency cell along all range bins for a specific motion. Lastly, the effectiveness of the proposed micro-Doppler signature enhancement strategy is investigated by means of onsite experiments and comparative classification. Both the feature enhanced spectrograms and classification results show that the proposed approach outperforms other state-of-art time-frequency analysis based feature extraction methods. The paper thus suggests the usage of an UWB radar system accompanied with the proposed range-max time-frequency representation (R-max TFR) approach for motion recognition and classification tasks in complex penetration scenarios such as through-wall, through-foliage, or for hand gesture related tasks such as human-computer interaction and intelligent home, for which the weak indistinguishable features prevail the spectrograms.

## I. INTRODUCTION

Nowadays, in response to the complex public security challenges (violent conflicts, terrorism, etc.) faced with our daily life and to avoid the mass casualties of combat forces on our side in modern urban warfare (obscured indoor scenes for street battle), intelligent surveillance of indoor human subjects' activities through opaque building walls becomes a very important research topic of great social and military importance and necessity. If penetrating detection and recognition of different indoor human activities could be realized, such operations like anti-terrorism operations, disaster rescue operations, and urban warfare operations would certainly experience a dramatic leap in available perception methods and tools for practitioners.

Generally, acoustic-based micro-Doppler analysis method is employed to perceive a specific type of subject motion, whether for human targets or non-human targets [1-7]. Whereas, several drawbacks of the approach restricted its wide application because of the intrinsic defects of the mechanical wave, which includes short detection distance, fast attenuation, opaque wall media penetration incapability, and low operating frequency resolution accompanied with low operating frequency.

Recently, encouraged by Dr. V.C. Chen's pioneering work on micro-Doppler effects in radar observations [8-12], radar-based micro-Doppler analysis technique has received extensive examinations and stands out as a much more effective sensing methodology than traditional acoustic-based method when detecting and identifying a specific subject in motion or its embedded movement pattern. To be specific, the periodic movement of a constituent part of the target, the limbs and torso for a human target, will introduce additional Doppler-shift to the backscattered radar echo signal. Since different moving body parts of the same target and the same movement pattern of different targets would all generate their own distinctive micro-Doppler signatures, this information could thus be

Qiang An* and Shuoguang Wang* contribute equally to this paper. Shiyong Li† and Ahmad Hoorfar† are the Co-corresponding authors for this paper.

Qiang An, Hao Lv, and Jianqi Wang are with the department of Biomedical Engineering, Fourth Military Medical University, Xi'an 710032, China (qiang.an.903@outlook.com, fmmulvhao@fmmu.edu.cn, wangjq@fmmu.edu.cn).

Shuoguang Wang and Shiyong Li are with the Beijing Key Laboratory of Millimeter Wave and Terahertz Technology, Beijing Institute of Technology, Beijing 100081, China (shuoguangwang@outlook.com, lisy_98@bit.edu.cn).

Ahmad Hoorfar and Wenji Zhang are with the Antenna Research Laboratory, Center for Advanced Communications, Villanova University, Villanova, PA 19085, USA (ahoorfar@villanova.edu, wenjizhang@gmail.com).

utilized for tasks such as target recognition, motion classification, etc.

Radar based micro-Doppler analysis has been successfully applied in such areas like small unmanned aerial vehicle (UAV) identification for flight safety screening at airports [13-15], through-foliage human target detection for low maintenance border control [16-20], pedestrian recognition in automatic driving [21, 22], smart home for intelligent life [23-25], human and animal discrimination for reduction of false alarm in surveillance and search-and-rescue operations [26-31], and hand gesture recognition for human-computer interaction [32-35], etc. These works have greatly promoted the development of the technology in intelligent sensing and estimation of the target's attribute via its intrinsic local or global movement patterns.

When it comes to indoor human activity detection and monitoring, tremendous work have investigated the problem from various aspects such as armed/unarmed personnel classification for public security purpose [36], human gait analysis for rehabilitation engineering and assisted living [37-41], human gait analysis for personnel's biometric identification [42], human gait analysis for early prevention of brain related disease [43, 44], and fall motion detection for elderly care [45-49]. However, these works mainly focused on the free space scenario.

For through-wall indoor human activity detection application, as with the introduction of the wall media, the electromagnetic (EM) wave experiences complex interactions with the opaque wall media (direct reflection from the front and rear of the wall, multiple reflections inside the wall, EM wave blockage and attenuation when propagating through the wall, and the associated dispersion and phase distortion, etc.). These factors would surely influence and even obscure the micro-Doppler features of behind wall indoor human activities.

Apart from that, ambient noises and rich multipath components in the real indoor environment further degenerated the obtained micro-Doppler signatures of behind wall indoor human activities.

Only limited literature was reported to deal with the through-wall indoor human activities detection problem. The work in [50] and [51] employed respectively a 6.5 GHz and a 5.8 GHz continuous wave (CW) radar system to detect and analyze the micro-Doppler signatures of different behind wall indoor human activities. The authors in [50] found that only the magnitude response of the spectrogram would be affected by the attenuation of the wall, while the work in [51] proposed that a moving target indication (MTI) filtering should be first applied to pre-process the range time intensity (RTI) plot so as to gain a significant improvement of signal-to-noise ratio (SNR) and micro-Doppler signatures, which process is vital to successfully classify different motions. However, both works provide no in-depth explanation and understanding as to the mechanism of the wall's effects and ways to mitigate such effects.

To figure out exactly what impacts would the wall media bring to the micro-Doppler signatures of behind wall indoor human activities, the work in [52] and [53] investigated the problem via numerical simulation and theoretical analysis, and they concluded that the constant phase term and the SNR deterioration of the micro-Doppler signatures are the two major effects caused by the introduction of the wall.

In a practical through-wall indoor motion detection scenario, the constant phase term is primarily composed of mutual coupling of the transceiving antenna pairs, direct reflection from the front and rear surface of the wall, and multiple reflections occurred inside the wall media. Since the range of these components remains constant in radar echo data, they don't exhibit time-varying phase shift. The term would present as strong near-zero direct current (DC) component in the time-frequency (TF) map, or in other words spectrogram, which component could overwhelm the already very weak micro-Doppler signatures of different human activities behind wall. While the SNR decrease of the micro-Doppler signatures is mainly caused by the wall blockage and attenuation of the EM wave. The two factors function simultaneously and make the subsequent motion recognition and classification extremely difficult.

There have been some reported literatures that use Generative Adversarial Network (GAN) to suppress the noises that spread the whole TF map so as to increase the SNR of the TF map [54], [55], and excellent denoising performance is achieved. Thus, denoising of the spectrogram is not of research interest of this work. Besides, we will point out later in this work that SNR deterioration effect is generally observable in a narrow band CW radar system. By utilizing an ultra-wideband signal, SNR performance of the obtained micro-Doppler signatures can be significantly improved using the proposed method in this work.

Special attention should be paid here since the micro-Doppler signatures of the body torso of a human subject in motion would also show up as zero (for in-place motions, like sitting, fetching, bending, etc.) or near-zero (for cross-place motions, like walking, crawling, etc.) DC component in TF map, even if the component is much weaker than its counterpart introduced by the wall effects. As a result, both components would prevail the obtained micro-Doppler signatures, which increases the difficulty of motion recognition and classification.

In the aim of removing the above discussed DC components so as to increase the contrast of TF map and enhance the high-frequency end characteristic features of TF map, the work in [56] proposed to deduct the near-zero frequency DC component directly in data matrix of TF map. However, the approach would lead to the interruption and discontinuity of the micro-Doppler signatures, which made it impossible to extract the corresponding micro-Doppler trajectories of different body segments in TF map, not to mention further understanding of the subtle embedded biometric or physiological information delivered via these trajectories. Thus, novel approach is still desired to mitigate such DC component in the spectrogram.

Actually, except for the wall's influence, the Doppler resolution degeneration along with the decrease of the working frequency is another factor that restricts the practical through-wall human motion detection work to move toward deeper and wider. Besides, the works in [50-53] all utilized TF map to identify and classify the behind wall human motions. In fact, apart from the time-frequency representation, there exists another rather important feature representation form, range over slow-time variation information, in which rich, detailed behind wall human body segments' movement information can be directly revealed provided that the range resolution is high enough. However, the above reviewed work all failed to make full use of it due to the limitation of the hardware platform. A narrow-band CW radar cannot provide enough range discriminative information. Even though a range-time profile was provided in [51], a bandwidth of 83.5 MHz could only achieve a range resolution of 3.6 m, which is far from enough to identify behind wall motions, especially for in-place motions which only span a small scope of range bin cells.

Thus, acquisition of as complete motion information as possible and high-resolution motion feature representation become two priority issues for performance improvement.

In order to exploit the range over slow-time variation information, many researchers turned to ultra-wideband (UWB) radar system. The work in [57] utilized an UWB impulse radar when analyzing the free space human gait variation information. The arms swings, the torso and legs motions can be clearly sighted in both range map and time-frequency map. Then, by employing a similar UWB pulse-Doppler radar, the work in [58], for the first time, proposed to jointly represent the micro-Doppler signatures and the high-resolution range signatures of free space indoor human motions in tri-domain, namely range, frequency, and time. The approach was referred to as joint range-time-frequency representation (JRTFR). In this way, a three-dimensional (3-D) data cube is constructed [59-61], in which both highly accurate range and micro-Doppler information can be simultaneously observed. Therefore, for through-wall indoor human motion detection application, an UWB radar system should be also embraced to capture as much motion information as possible.

As long as a JRTFR data cube is available, the subsequent motion recognition and classification can be done either directly in tri-domain [62-64] or in its two-dimensional (2-D) subset feature slice domain, for example time-integrated range-Doppler domain [65-68], time-frequency domain [69], or range-time domain [70]. However, the 2-D subset feature slice representation could only depict a specific motion under investigation from a unilateral point of view. The information loss accompanied with the absence of the third dimension is bound to affect the follow-up motion recognition and classification. Then, to avoid the above shortcoming and improve the motion classification performance, the work in [71, 72] proposed to fuse the motion information extracted from different 2-D subset feature representation domains. Since mutually complementary motion information of these domains is taken into consider unitarily, the approach could thus yield higher motion classification accuracy when compared to the case that only a single feature domain is utilized. Yet, the above domain level fusion (feature fusion) approach still didn't fully exploit the internal dependency and correlations of a 3-D JRTFR data cube. Only two or three domain slices or domain projections were employed. To comprehensively consider the underlying inter-dependency and cross-correlations among all three radar signal variables, the work in [62-64] attempted to directly process the 3-D data cube. The method could achieve extremely high classification accuracy. Whereas, the e-CLEAN algorithm [73] should be first applied to preprocess the data cube for the purpose of suppressing the unwanted distortions and noise components inside the 3-D data structure. Moreover, the multi-dimensional principal component analysis (MPCA) method [74] should be carried out to realize the unsupervised multi-linear feature extraction and classification. Both procedures in nature require tentative iterative operations, which render the approach computational cost and inapplicable of real-time processing.

Therefore, to simplify the behind wall motion recognition and classification process, meanwhile preserve the most significant internal dependency and correlations of the original 3-D data cube, proper dimensionality reduction inspired motion feature enhancement approach, rather than simple projection or slicing, becomes imperative and extremely important.

The work in [66] described the typical time-frequency analysis method to process the ultra-wideband signal. The slow-time radar returns expanded in different range bins are first added up together coherently, then the short-time Fourier transform (STFT) is applied to the above range accumulated slow-time signal. In this way, body segments' movement information spread along different range bins is considered. For the convenience of reference below, the approach can be named as RA-STFT. As can be anticipated, when the method is used to process the UWB radar returns of a real behind wall human motion, the obtained micro-Doppler signatures would present rather faint characteristic features, meanwhile it would be filled with noise components. Several factors contribute to this result just as analyzed above. Firstly, the wall effects kick in here and accounts partially for the degeneration of the micro-Doppler signatures. Secondly, to acquire enough penetration ability, a low center-frequency UWB radar is often employed, which is followed by a relatively low Doppler resolution. Thirdly, the near-zero DC component introduced by the body torso is not removed in TF map and preserves to be strong. The component would surely conceal the high-frequency features of the limb's motion, which mainly served as characteristic features for motion recognition and classification. Lastly, for each range bin sample, the body torso's movement possesses the most significant energy, while the motions of head, arms, and legs spread across the range bins and weaken along the range direction due to their small radar cross section (RCS). Range accumulation mixed these components up together. The practice would only make the micro-Doppler signature of the torso motion being strengthened while that of the limb motions being attenuated.

Instead of the approach of RA-STFT, the work in [75] proposed to first calculate the TF map of each slow-time signal and then weigh these TF map frames using their corresponding inverse energy coefficients. By applying a large weight to the TF map frame with weak energy, and a small weight to the TF map frame with strong energy, the weighted TF map of behind wall human motions could thus possess much more significant and distinct micro-Doppler features, especially for signatures at high-frequency end. The approach is termed as comprehensive range accumulation time-frequency transform (CRATFR). Whereas, as indicated in the obtained spectrograms in [75], the lower frequency components of the TF map would be enhanced more remarkable than the higher frequency components, which reflects in spectrogram as strong near-zero DC clutters. Meanwhile, heavy background clutters prevail the spectrogram, which is the side effect of the noise components when weighting and superimposing the TF map frames.

In this paper, we put forward a novel strategy to overcome the drawbacks of the above reviewed feature enhancement approaches. Firstly, a stepped-frequency continuous wave (SFCW) multiple-input multiple-output (MIMO) UWB radar is utilized to collect the raw range map of behind wall human motions. Secondly, a high-pass filter is applied to remove the wall related reflections. Thirdly, a JRTFR data cube is constructed by applying STFT successively to every slow-time instance. Then, inspired by the fact that behind wall human motions generally spread out over time in range map, and no two segments of the same motion would be found to be overlapped when observing the range map along the range direction for each fixed time instance, which means that range maximization, rather that projection (integration), or slicing, would capture the most significant motion information at that exact time slice. Thus, this work proposes to extract the maximum value of each TF cell along the range dimension in the 3-D data volume. In this way, a range-max enhanced micro-Doppler signature is obtained. Such a maximization strategy could not only effectively compress the original high dimension radar data cube (RDC) but also preserve the most significant and intrinsic range diversity characteristics of behind wall motions. We call the approach range-max time-frequency representation (R-max TFR). At last, onsite experiments are carried out to verify the performance of R-max TFR. Experimental results show that the approach outperforms both RA-STFT and CRATFR.

The paper is organized as follows. In Section 2, the traditional through-wall indoor human motion micro-Doppler analysis approach and wall effects mitigation method are introduced. In Section 3, the proposed R-max TFR method is described in detail. In Section 4, the experimental comparisons of the proposed method with the above reviewed other two methods are presented. In Section 5, the concluding remarks are provided.

## II. THROUGH-WALL HUMAN MOTION MICRO-DOPPLER ANALYSIS AND WALL EFFECTS MITIGATION METHOD

### A. Signal model for SFCW UWB radar system

Consider a typical through-wall indoor human activity detection geometry as depicted in Fig. 1. A human subject walks back and forth behind a single layer of a concrete wall. An antenna array operating in a multi-static mode probes the target space of interest. Assuming the transmitted SFCW UWB signal contains $N$ frequency bins in total, the $n$-th sweeping

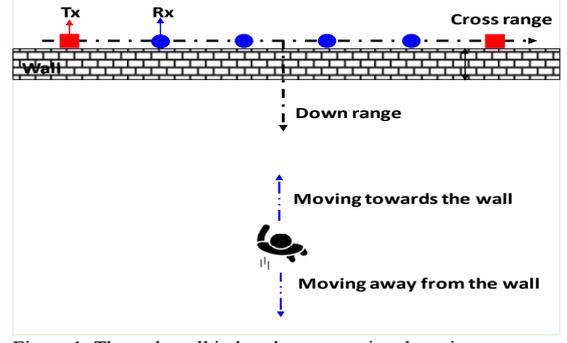

Figure 1: Through-wall indoor human motion detection geometry

frequency is expressed as follows,

$$s_t(t) = rect\left(\frac{t}{T}\right) exp\left(j2\pi\left(f_0 + n\Delta f\right)t\right) \tag{1}$$

where $rect$ denotes the window function, $T$ is the modulation period, $f_0$ is the starting frequency, and $\Delta f$ refers to the frequency step.

Then, the backscattered radar echo signal modulated by the moving human target behind the wall can be written as,

$$s_r(t) = a \cdot rect\left(\frac{t-t_d}{T}\right) exp\left[j2\pi\left(f_0 + n\Delta f\right)\left(t-t_d\right)\right] \tag{2}$$

where $a$ stands for the reflectivity coefficients of different body parts in motion, which value is relatively small due to the constrained observable body parts' cross-section in the radar line-of-sight. $t_d = 2R(t)/c$ denotes the instantaneous time-delay information of a particular moving body part at distance $R(t)$. In addition, the direct reflection components from the front and rear surface of the wall, and the internal multiple reflection components all contribute to $a$ and $t_d$. However, just as discussed above, these components would reflect as strong wall clutter components in range map and as constant phase terms (strong DC components) in TF map.

After demodulation operation, the received signal can be represented as below,

$$s_{if}(t) = a \cdot rect\left(\frac{t-t_d}{T}\right) exp\left[-j2\pi\left(f_0 + n\Delta f\right)t_d\right] \tag{3}$$

The signal $s_{if}(t)$ contains all the valuable information regarding the behind wall human motions. With an additional discretization process, equation (3) can be further reformulated as follows,

$$s_{if}(n,m) = a \cdot exp\left[-j2\pi\left(f_0 + n\Delta f\right)t_d(n,m)\right] \tag{4}$$

The above discretized form of the received signal is typically organized in a 2-D matrix form, in which each row corresponds to the echo signals sampled using the very same probing frequency at different pulse repetition instance, while each column corresponds to the captured frequency response of the behind wall human motion using the above emitted SFCW UWB signal in equation (1). Then, by applying the inverse discrete Fourier transform (IDFT) to the collected frequency samples in each column, a range map (range v.s. slow time) is readily derived as below,

$$s(l,m) = IDFT_L(s_{if}(n,m)) \quad l = 1,2,\ldots,L \quad (5)$$

where $L$ dentoes the number of FFT, $m$ is the number of slow-time data samples decided by the data collection time.

The range map depicts human motion's range variation information over slow-time dimension. However, only cross-place motions, like walking, crawling and jogging, can be effectively recognized by extracting their moving trajectories in range map. One is faced with difficulty when handling with in-place motions such as sitting-down and falling, because such motions don't exhibit large range spans, which makes them unrecognizable by utilizing only the range maps. From feature representation point of view, the range map is thus inadequate to describe the subtle difference among those motions.

Fortunately, there exists another common and useful feature representation form, the STFT based micro-Doppler analysis, which could make up for the deficiency of the range map.

### B. Micro-Doppler analysis using STFT

Radar is capable of sensing and capturing the Doppler and micro-Doppler signatures related to moving targets (human or non-human targets), which is caused by rotating, oscillating, or vibrating of the constituent parts of the target. As for the human motions, Doppler refers to the frequency change caused by the motion of the body torso, while micro-Doppler refers to that caused by other body parts, such as foot and legs. The latter is centered around the main Doppler shift in TF map, serving as characteristic features for motion recognition and classification.

Time-frequency analysis has been extensively employed in various applications. In this paper, the STFT based time-frequency analysis is adopted to generate the spectrogram. It is mathematically expressed as follows,

$$SPEC(m,k) = \left| \sum_{t=0}^{K-1} s(m+t)h(t)exp(-j2\pi kt/K) \right|^2 \quad (6)$$

where $s(m)$ is the range accumulated radar data (summing up the slow-time signals of the range map along the range direction), and $h(t)$ is a window function used to truncate the time series data, whose length is decided by the particular task. While the length of the window should be carefully chosen to strike a balance between time resolution and frequency resolution. The time-frequency analysis method depicted in equation (6) is precisely the aforementioned RA-STFT.

If range accumulation is dropped and STFT is directly applied to each slow-time instance, then a series of TF maps can be obtained as below,

$$TFR(l,m,k) = \left| \sum_{\substack{l=0 \\ q=0}}^{\substack{K-1 \\ L-1}} s(l,m+q)h(q)exp(-j2\pi kq/K) \right|^2 \quad (7)$$

where $s(l,m)$ is the slow-time signal of $l\text{-}th$ range bin, $h(q)$ is the window function. By applying a weighing procedure as proposed in [75], the CRATFR spectrograms can then be calculated as follows,

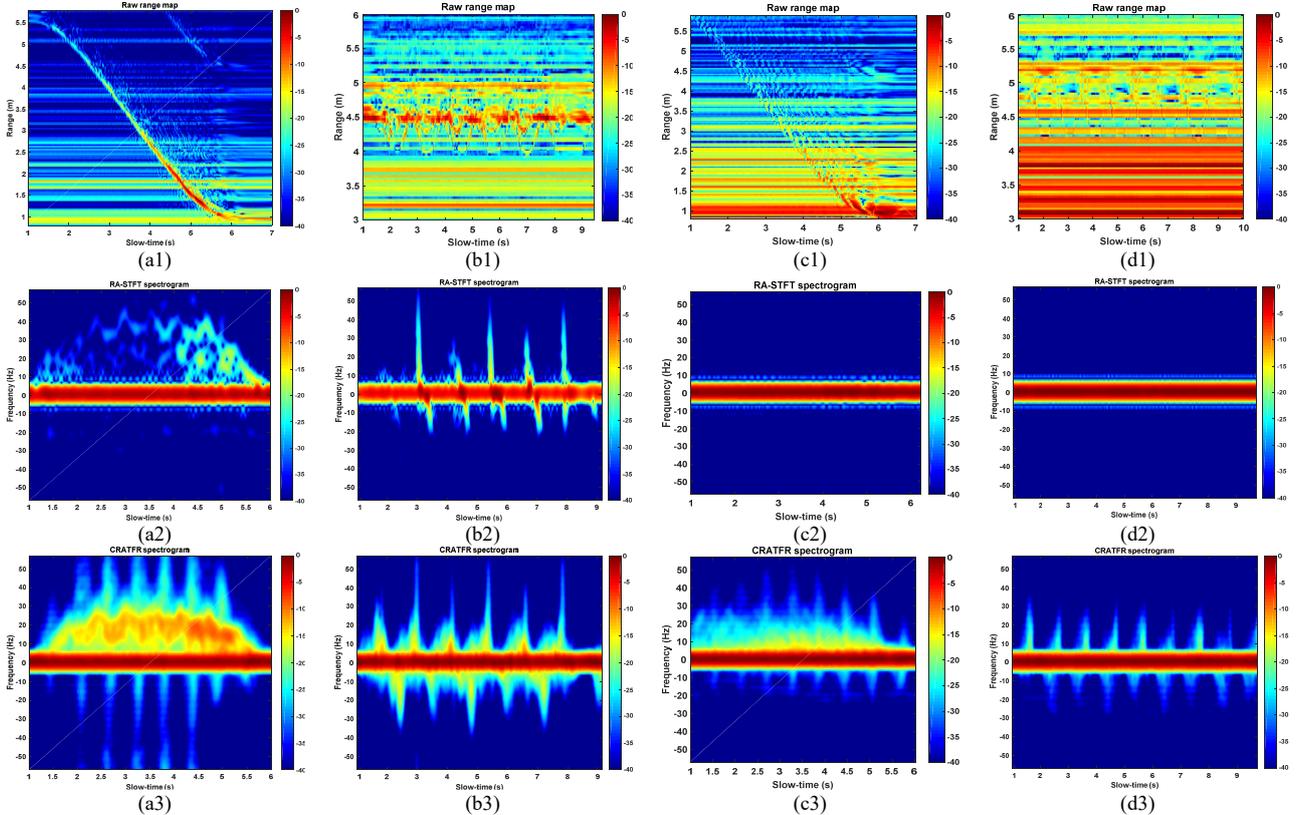

Figure 2: Raw range maps and corresponding spectrograms for two types of free-space and behind-wall motions, walking towards the wall (walking towards the radar front for free-space case) and in-place boxing. (a1) Range map for walking towards the radar front in free space. (b1) Range map for free-space in-place boxing. (a2) Spectrogram of (a1). (b2) Spectrogram of (b1). (a3) Range map for walking towards the wall. (b3) Range map for behine wall in-place boxing. (a4) Spectrogram of (a3). (b4) Spectrogram of (b3).

$$SPEC(m,k) = w_1 * TFR(1,m,k) + w_2 * TFR(2,m,k) + \quad (8)$$
$$... + w_L * TFR(L,m,k)$$

where $w$ is the weight coefficient as defined in [75].

For the experiments included in the following part of this paper, the RA-STFT spectrograms and CRATFR spectrograms are both generated using a Hanning window with a length of 32, and an overlap of 31 samples.

### C. Wall's influence on range map and spectrogram

For through-wall motion detection task, the influence of the wall is mainly manifested in two aspects: strong wall clutters in range map; near-zero DC components in TF map.

To intuitively illustrate the wall's influence on range map, we carried out comparative experiments using the SFCW UWB radar system depicted in Section 4. Two type of motions, walking towards the wall (for free space detection case, walking towards the radar front) and in-place boxing, are investigated. The detailed onsite measurement scene can be found in Section 4.

Fig 2 (a1), (b1), (c1), and (d1) show respectively the range maps for walking towards the wall (walking towards the radar front) and in-place boxing, collected in free space and through single layer of reinforced concrete wall scenarios. It can be clearly observed that the range maps in Fig. 2 (c1) and (d1) were severely overwhelmed by strong clutters when compared to the free space range maps in Fig. 2 (a1) and (b1), which is mainly caused by the wall reflections. Meanwhile, the trajectories of behind wall human motions is hardly recognizable in the range maps because of the wall blockage and attenuation of the EM wave.

A similar observation can be found in the obtained micro-Doppler signatures using RA-STFT. Fig. 2 (a2) and (b2) are the mico-Doppler signatures for walking towards the radar front and in-place boxing, respectively. Strong near-zero DC component appeared in both signatures, which is primarily induced by the modulation of the human body torso. In the meantime, Fig. 2 (c2) and (d2) show the micro-Doppler signatures of the same two motions in through-wall case. The near-zero DC clutters brought in by the wall reflections occlude completely the micro-Doppler signatures.

Thus, the wall clutters should be first suppressed before any further analysis been conducted.

### D. Proposed wall effects mitigation approach

In through-wall human subjects detection application and through-wall radar imaging application, mean value subtraction [76] and Singular Value Decomposition (SVD) based subspace projection [77] are two widely used methods to remove the wall clutters. Whereas, even though mean value subtraction method can mitigate stationary wall clutters effectively, its usage would bring in additional DC clutter components to time-range cells where there were supposed to be non-target motion regions. The factor affects cross-place motions most significantly, blurring the motion tracks, and meanwhile cluttering the micro-Doppler signatures. SVD-based subspace projection method, on the other hand, depends heavily on manual intervention. The number of singular values to be selected for signal space and wall clutter space decomposition, varies with different through-wall practices (different wall materials and diverse wall parameters). Moreover, the effectiveness of the approach remains problematic. In certain cases, it cannot remove the wall clutter thoroughly, while in other cases the wall clutter might be over-suppressed, even with the cost of suffering the valuable motion information to some extent. Other than that, SVD-based subspace projection method is hard to be implemented on a portable radar platform due to its computational cost and non-real-time processing defect.

In this work, to address the wall clutter removal issue, we suggest the use of a finite impulse response (FIR) high-pass filter (HPF). There are two reasons to use this filter. First, the FIR filter can ensure strict linear phase characteristics while designing arbitrary amplitude-frequency response, which means the phase information of behind wall motions can be well preserved when conducting wall mitigation using a FIR filter. However, the above two reviewed methods don't possess this feature. Another reason is that the HPF can be designed to keep very sharp decay within the transition band. Such a filter in effect allows only the signal components with frequencies larger than the cut-off frequency to pass and stops completely the components below the cut-off frequency. Since the wall clutters constitute mainly the DC and low frequency components of slow-time signals in range map, they can thus be effectively removed by applying such a filter for each slow-time signal along the range direction. The order of the proposed FIR HPF is selected to be 112 in this paper, while the cut-off frequency is set to be 3 Hz based on our on-site experiments.

Fig. 3 (a1), (a3), (a5), (b1), (b3), and (b5) show respectively the wall clutter removed range maps using mean value subtraction method, SVD-based subspace decomposition method, and high-pass filtering method for two behind wall motions, walking towards the wall and in-place boxing. One thing to note is that only the largest singular value component is thought to account for the wall reflections in SVD decomposition method of this paper because the wall clutters possess the most significant energy in range map. It was thus projected to wall clutter space. As can be observed in Fig. 3 (a1) and (b1), additional DC clutters were induced around the motion trajectories by averaging procedures, even though the wall clutters are removed using mean value subtraction method. From subspace decomposed results in Fig. 3 (a3) and (b3), it can be seen that the wall clutters are not completely suppressed and the motion trajectories is partially blocked by residual wall clutter, which means that the wall space should be composed by not only the largest singular value but also the second largest or even more singular values. In practice, the wall removal performance of the approach lies in how to conduct the most appropriate decomposition to achieve a complete separation of clutter space and motion signal space. By comparison, the FIR HPF filtered range maps in Fig. 3 (a5) and (b5) exhibit the best wall removal performance and simultaneously preserve the most authentic and complete trajectories for two motions.

In terms of the corresponding spectrograms in Fig. 3 (a2), (b2), (a4), (b4), (a6), and (b6), more distinct micro-Doppler signatures with much weaker DC components are observed as compared to the results in Fig. 2 (a4) and (b4). However, near-zero DC components still exist in mean value subtracted micro-Doppler signatures in Fig. 3 (a2) and (b2) and in subspace projected micro-Doppler signatures in Fig. 3 (a4) and (b4). The former is caused by the induced additional DC clutters surrounding the real motion trajectories, while the latter is the result of incomplete removal of wall clutters. Meanwhile, the spectrograms in Fig. 3 (a6) and (b6) present the least significant near-zero DC components as expected, which in effect illustrates the effectiveness of the proposed wall clutter removal method from another point of view. Thus, the motion recognition and classification task could attain an improved accuracy with the enhanced micro-Doppler signatures. Note, the remaining near-zero DC component in the spectrograms is primarily attributed to the motions of body torso, which appears in range maps as the trunk of motion trajectories. Since the component exists in the entire motion process and occupies generally multiple range bins, exhibiting irregularity in both time and range dimension, it cannot be mitigated using any of the above reported wall removal methods. One last thing to note is that RA-STFT method is applied when obtaining these spectrograms.

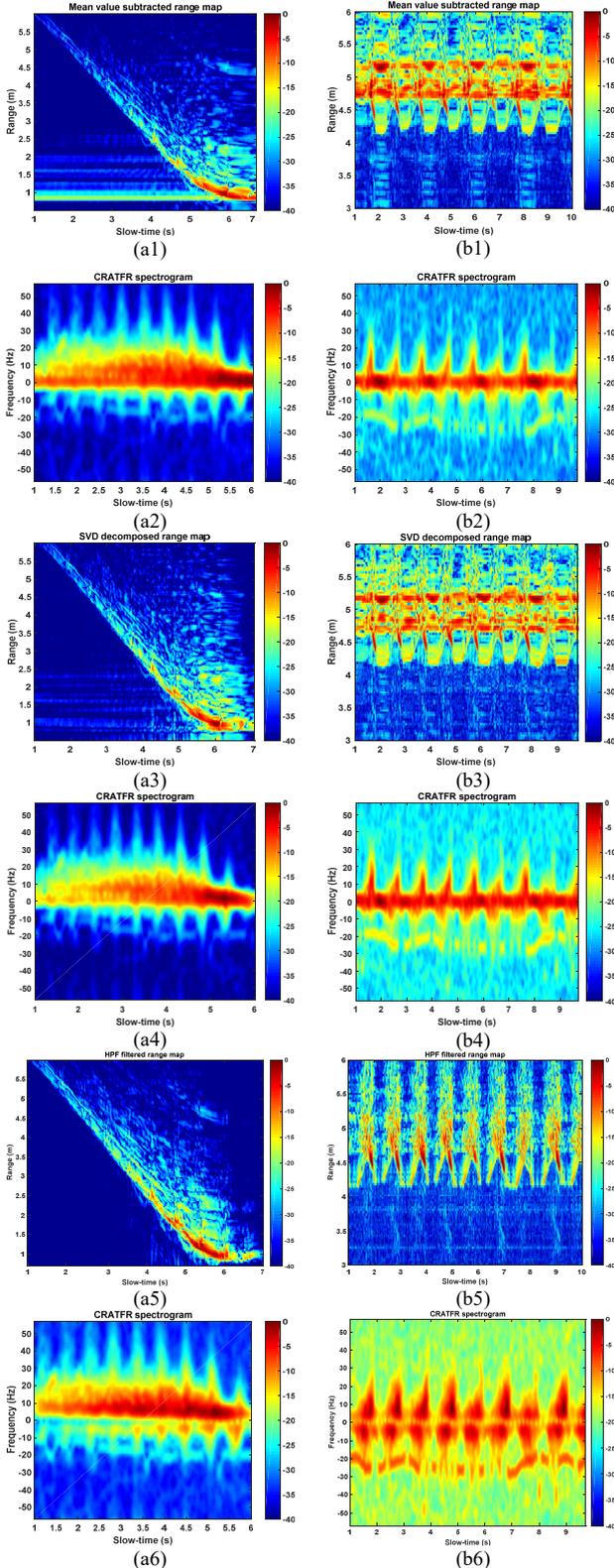

Figure 3: Mean value subtracted range maps, SVD decomposited range maps, high-pass filtered range maps, and corresponding spectrograms for two types of behind-wall motions, walking towards the wall and boxing. (a1), (b1) Mean value subtracted range maps for walking towards the wall and boxing. (a2), (b2) Spectrograms of (a1) and (b1). (a3), (b3) SVD decomposited range maps for walking towards the wall and boxing. (a4), (b4) Spectrograms of (a3) and (b3). (a5), (b5) High-pass filtered range maps for walking towards the wall and boxing. (a6), (b6) Spectrograms of a(5) and (b5).

Whereas, even if the wall is removed with the aid of high-pass filtering, the classification of behind wall in-place motions is still faced with difficulty due to their subtle indistinguishable differences of range maps or spectrograms. The work in [78] and [79] adopted a multi-stage classification strategy to increase the classification accuracy. The strategy is encouraging. However, the permutation and combination of classifiers is only capable of enhancing the classification performance, while the recognition of a specific motion type remains unimproved. Thus, more distinct characteristic features are still desired to discriminate the inter-class in-place motions. Fortunately, with the employment of an UWB radar system, we could dive into this problem and go a step further by exploiting the available UWB information.

## III. RANGE-MAX TIME-FREQUENCY REPRESENTATION

In this work, a SFCW UWB radar system is utilized to collect the raw echo data of behind wall human motions. As long as a range map is obtained, the subsequent motion feature representation can be extended to either time-frequency domain or time-integrated range-Doppler domain as described above. Whereas, the above three types of feature representations could only perceive and comprehend the behind wall motion features partially and incompletely. In order to obtain a global insight into the inter-dependence and correlation relationships among three radar variables (time, range, frequency), the data cube analysis is conducted by applying STFT to each slow time instances along range bins of the range map. It mathematically processes the same form as equation (7) as follows,

$$RDC(l,m,k) = TFR(l,m,k) \quad (8)$$

The generation process of an RDC is illustrated as below in Fig. 4. In this way, the range, time, and frequency information of a specific behind wall human motion is highlighted and jointly considered in a volumetric cubic form. However, the computational cost accompanied with the high-dimensional feature extraction (including the noise suppression procedure using clean algorithm) and classification makes it inapplicable for real time processing. Methods like RA-STFT and CRATFR were thus proposed to meet the challenge while making full use of the UWB information of behind wall motions, whereas, the faint characteristic motion features and strong background clutters in the obtained spectrograms limited the methods' capability of improving the motion recognition and classification performance. Therefore, from the motion feature representation perspective, novel approach is still needed to achieve a more distinct representation of the behind wall motions.

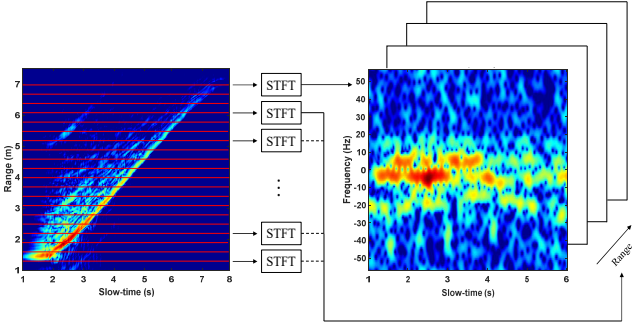

Figure 4: Radar data cube generation using STFT processing

In this paper, we propose a novel form of feature representation by compressing the RDC in equation (8) using a maximization strategy. It first emulates every TF cell along the range direction of the obtained radar data cube and then extract the cell with the strongest energy representation. In this way, a new type of feature representation form is arrived. Mathematically, the process can be represented as follows,

$$TFR_{Rmax}(m,k) = \max_{l}\left[ RDC(1,m,k) \right] \qquad (9)$$

The approach is termed as range-max time-frequency representation (R-max TFR). Accordingly, the generated spectrogram is called R-max spectrogram. Our proposed approach conducts in nature a type of dimensionality reduction. When compared to the above discussed integrated projection or slicing, which either enhance the noise and low-frequency components or cannot depict the overall features of a specific motion, R-max TFR could not only extract the most intrinsic motion signatures in the original RDC but also suppress the unwanted noise components in RDC. Thus, enhanced features could be expected to obtained using R-max TFR.

The theoretical rationality of the effectiveness of the approach rest with two facts. The first one is that a specific human motion typically spans a certain consecutive number of range bins when observing from any fixed slow-time snapshot, in which the motions of the limbs occupy mainly the two sides of the range bins, while the motions of the body torso compose the central part. By applying STFT to each slow-time instance in that range bins, a series of spectrograms can be obtained. Each spectrogram only reflects the instantaneous motion signatures when the movement is conducted in that fixed range distance. The other fact is that the motion trajectories do not overlap for any fixed time instance as with the extension of the observation process. Therefore, if we zoom into the above obtained spectrogram series and analyze the motion signatures around a specific fixed time instance, the lower-frequency components standing for the motions of torso will be highlighted only in a small number of spectrogram slices, while the same conclusion holds true for the higher frequency characteristic signatures of the limb motions. Thus, for each TF cell, maximization along range direction in above the spectrogram series (RDC) rather than integration or slicing, could not only capture the most authentic motion signatures with high frequency features highlighted and enhanced but also suppress unwanted noises. As is well acknowledged, only the motion signatures of the limbs serve as characteristic features for motion recognition and classification. The generated R-max spectrograms could thus improve the classification accuracy of different indoor human motions.

In fact, a similar procedure can be adopted to produce a time-max range-Doppler representation (T-max RDR), and thus a T-max RD signature is obtained. Whereas, because the size of a real indoor space is often constrained and the Doppler frequencies of different indoor motions don't vary significantly, the T-max RD is not suitable to analyze the behind wall human motions.

Several advantages are accompanied with the proposed R-max TFR approach. Firstly, complex clutter and noise components suppression method is avoided. The motion detection task can be conducted with high efficiency. Secondly, a spectrogram with enhanced motion characteristic features is obtained, which is rather useful to meet the challenge for motion recognition and classification tasks where only a low operating frequency is available, or the detection scenario is complex. Lastly, when compared with the state-of-art micro-Doppler analysis method, CRATFR, the spectrograms obtained using R-max TFR approach exhibit much more evident and distinct motion features.

In order to examine the feature enhancement performance of the proposed R-max TFR method, experimental comparisons are carried out for two types of behind wall motions: walking towards the wall and in-place boxing. The experimental scenarios and setup keep unchanged with that of the case when collecting the raw range map in Fig. 2. CRATFR method and R-max TFR method are applied to the collected raw range maps of the two behind wall motions.

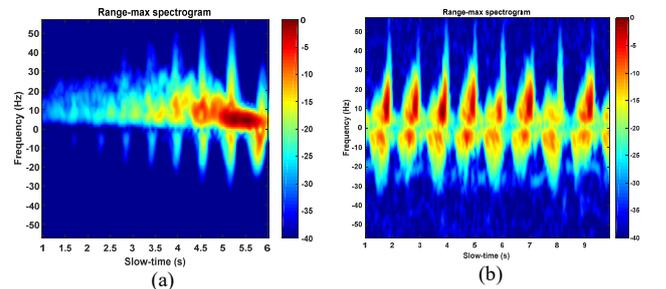

Figure 5: Spectrograms calculated using R-max TFR for two types of behind wall motions, walking towards the wall and in-place boxing. (a) R-max spectrogram for walking towards the wall.. (b) R-max spectrogram for in-place boxing.

Fig. 3 (a6) and (b6) show respectively the spectrograms for walking towards the wall motion and in-place boxing motion calculated using CRATFR method. Obvious background clutters spread the whole micro-Doppler signatures, which is the result of the accumulation of noise components. There is no doubt such spectrograms could hardly produce an optimistic classification accuracy. Fig. 5 (a2) and (b2) display the spectrograms of the two motions obtained using R-max TFR method. Much clean spectrograms are obtained, in which not only the background clutter is removed but also the characteristic features at high frequency end become more prominent. The results prove for certain the excellent performance of our proposed feature enhancement approach.

## IV. EXPERIMENT RESULTS

To further verify the effectiveness of the proposed feature enhancement approach, a series of through wall indoor human motion detection experiments were conducted. A low center-frequency SFCW UWB radar developed by our group [80], consisting of 2 transmitters and 4 receivers, is employed to detect the behind wall motions. The detailed parameters of the radar system show as follows: the operating frequency ranges from 380 MHz to 4.4 GHz, the frequency step is 5 MHz, and the pulse repetition frequency (PRF) is 113 Hz. Owing to the low operating frequencies and the large bandwidth, the system possesses both excellent opaque media penetration capability and high range discrimination ability.

The experimental scene is as shown in Fig. 6. Fig. 6 (a) is the free space motion detection scenario, while Fig. 6 (b) depicts the through wall motion detection scenario, in which the antenna array is placed against the exterior surface of a single layer of reinforced concrete wall. The wall is about 30 cm thick. A human target stands 5 m from the radar front performing different type of motions. Through wall field data of human motions is collected utilizing the above UWB radar system.

To be specific, we study 10 types of common indoor human motions in total, including forward walking, backward walking, sitting on a chair, standing up, picking things up, forward crawling, backward crawling, forward falling, backward falling, and in-place boxing. The illustration of these motions can be found below in Fig. 7. Since we have already compared and discussed the feature enhancement performance of the proposed R-max TFR approach when handling with forward walking motion and in-place boxing motion, we then continue to analyze whether the approach remain to be effective when applied to other types of motions.

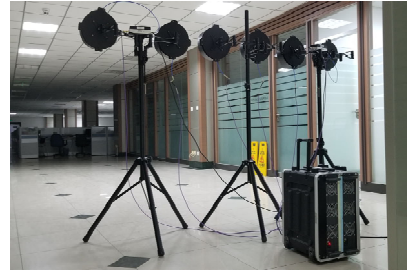

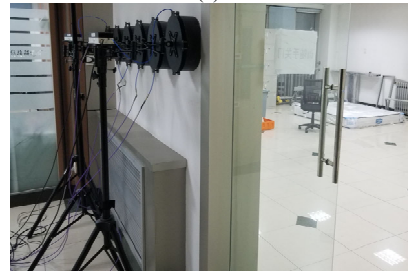

Figure 6: Indoor human motion detection geometry. (a) Free space measurement setup. (b) Through single layer of reinforced concrete wall measurement setup.

Furthermore, a behind wall motion dataset is built to facilitate the subsequent motion classification. 10 volunteers (3 females and 7 males) were recruited to conduct the above 10 types of indoor motions. Each motion lasts for 30 seconds. Every test subject repeats each motion for 10 times. Thus, 100

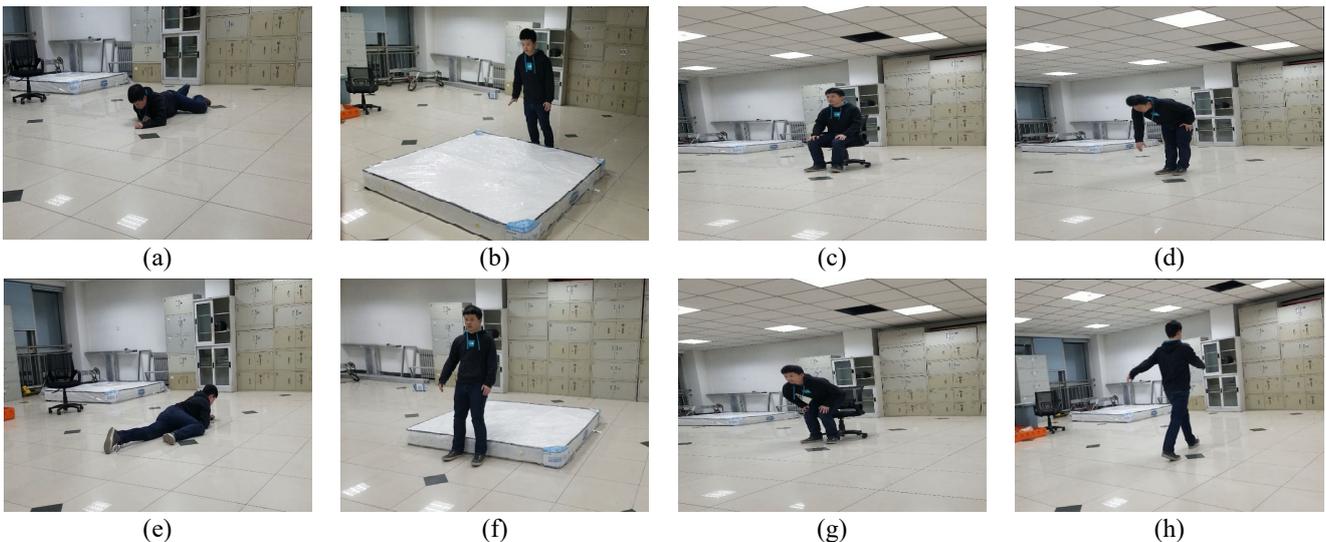

Figure. 7: 8 types of common behind wall indoor human motions. (a) Backward walking. (b) Sitting down. (c) Standing up from seat. (d) Fetching down and up. (e) Forward crawling. (f) Backward crawling. (g) Forward falling. (h) Backward falling.

samples are obtained for each motion in the dataset.

It should be noted that forward means moving towards the wall and vice versa. In order to validate the performance of the proposed approach, a comparative study with the state-of-art CRATFR method is conducted. One more thing to note, the FIR HPF should be first applied to remove the wall effect before any further feature representation is carried out.

### A. Motion feature enhancement using proposed T-max TFR

Fig. 8 shows row by row the raw range maps, HPF filtered range maps, CRATFR spectrograms, and R-max spectrograms for forward walking motion, sitting down motion, standing up from seat motion, and fetching down and up motion.

As can be observed in the first row of Fig. 8, motion trajectories are totally overwhelmed by strong wall clutters. When FIR HPF is applied to the raw range maps in Fig. 8

(a1-d1), the wall clutter is significantly suppressed and motion trajectories are highlighted. Since we previously concluded through comparison that CRATFR outperforms RA-STFT in that motion features at high frequency end can be enhanced more prominently using CRATFR method. Thus, CRATFR method is used to process the wall clutter removed range maps in the second row of Fig. 8. It can be seen in the obtained CRATFR spectrograms in the third row of Fig. 8 that even if the characteristic motion features are strengthened, the background noises clutter the CRATFR spectrograms, while the R-max spectrograms don't possess this defect. In fact, clean feature maps are obtained using R-max TFR approach, meanwhile the characteristic features at the high frequency end of the spectrograms, which represents the motion features of the limbs, are highlighted much more intensively. Both factors

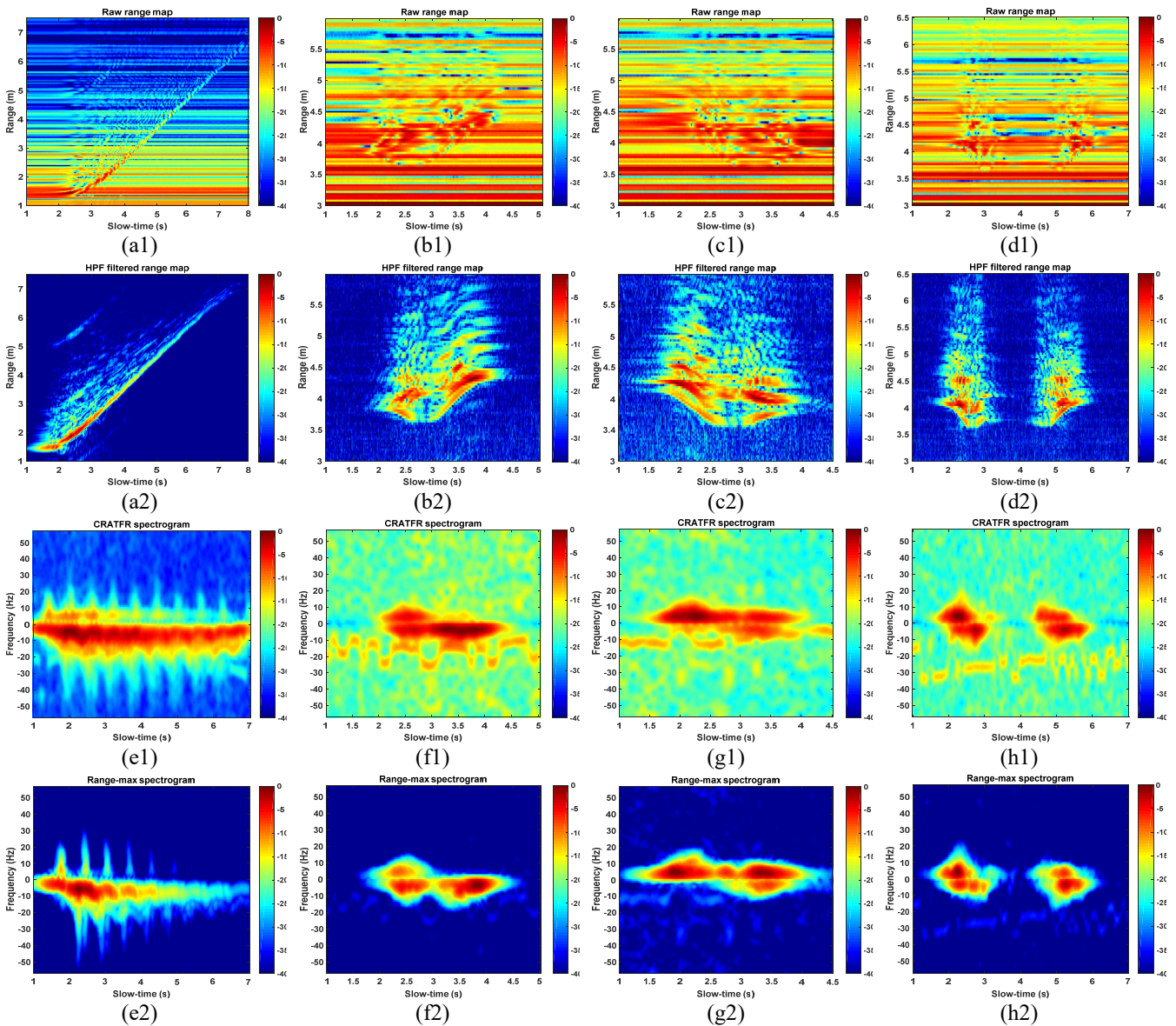

Figure 8: Raw range maps, HPF filtered range maps, CRATFR spectrograms, and R-max spectrograms for back ward walking motion, sitting down motion, standing up from seat motion, and fetching down and up motion. (a1-d1) Raw range maps for the four behind wall motions. (a2-d2) HPF filtered range maps for the four behind wall motions. (e1-h1) CRATFR spectrograms for the four behind wall motions. (e2-h2) R-max spectrograms for the four behind wall motions.

favor both motion recognition and classification tasks.

For forward crawling motion, backward crawling motion, forward falling motion, and backward falling motion, their raw range maps, HPF filtered range maps, CRATFR spectrograms, and R-max spectrograms can be found in Fig. 9. A similar conclusion can be drawn by observing the motion features in spectrograms calculated using R-max TFR method.

Thus, it can be concluded with confidence that a spectrogram possessing enhanced motion features can be obtained by utilizing the R-max TFR approach proposed in this work. The approach could aid not only the motion recognition task but also the subsequent motion classification task.

### B. Classification results

In this part, we dive into the classification task and to investigate whether the novel R-max spectrograms could increase the classification accuracy of the above 10 types of behind wall motions.

2D-PCA method is first applied to realize feature extraction [81]. The k-Nearest Neighbors (kNN) classifier is then employed to discriminate different behind wall motions [82]. The value for k is set to be 3 in this work.

The spectrograms calculated using the proposed R-max TFR method for 10 types of behind wall motions are firstly converted to grayscale images. Then, they serve as the inputs of Singular Value Decomposition (SVD). The training procedure is similar to that typically employed in face recognition task. 80% of the dataset is employed to train the classifier, while the is used as testing. The training and testing sets are selected randomly. Meanwhile, 1000 Monte Carlo trials are applied to eliminate the random selection effect.

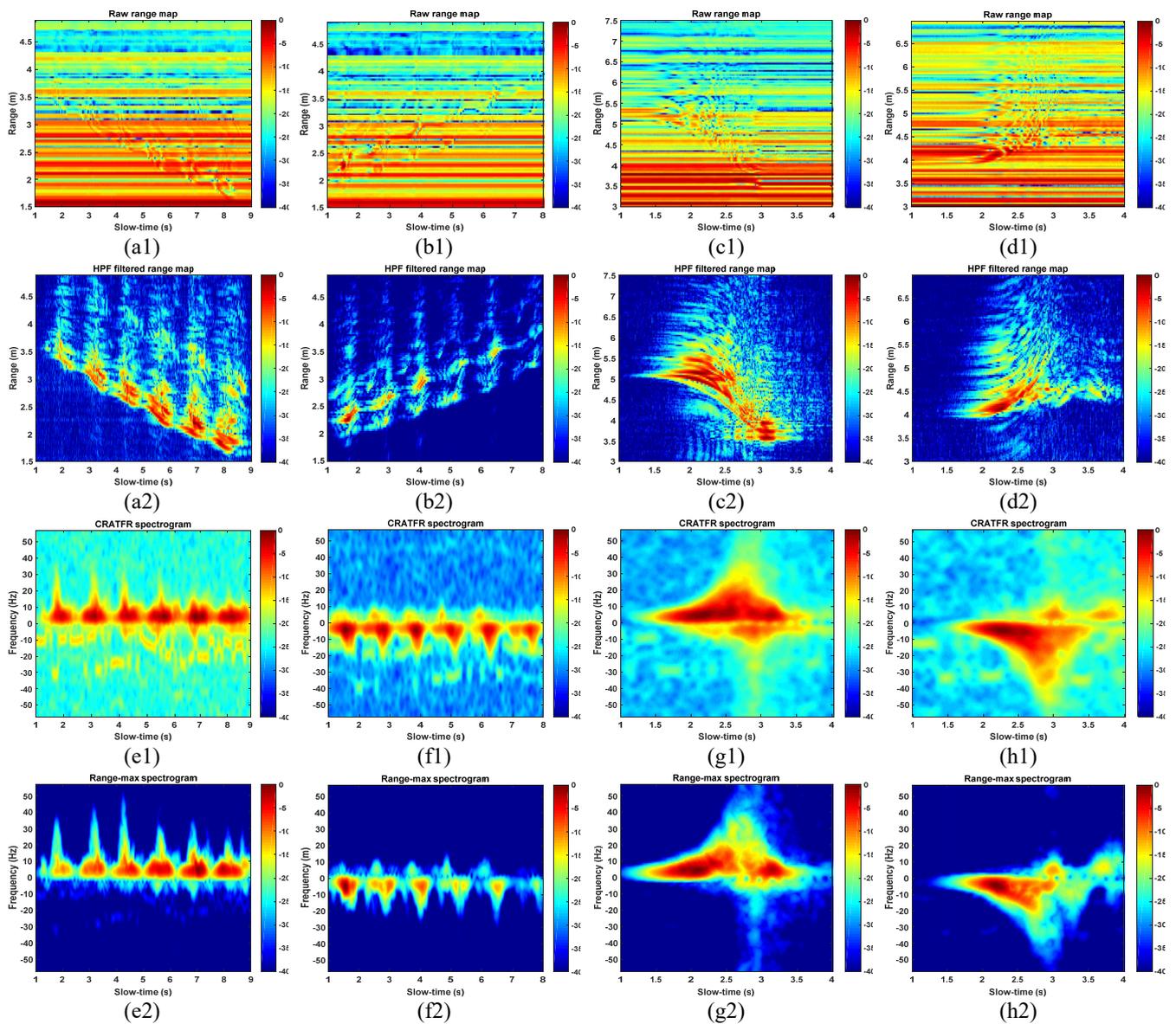

Figure. 9: Raw range maps, HPF filtered range maps, CRATFR spectrograms, and R-max spectrograms for forward crawling motion, backward crawling motion, forward falling motion, and backward falling motion. (a1-d1) Raw range maps for the four behind wall motions. (a2-d2) HPF filtered range maps for the four behind wall motions. (e1-h1) CRATFR spectrograms for the four behind wall motions. (e2-h2) R-max spectrograms for the four behind wall motions.

The resulted confusion matrix for test datasets is given as below in Table 1. The rows of the confusion matrix represent the actual motion classes, while the columns denote the classes predicted by our classifier. It can be concluded in Table 1 that through wall indoor human motion classification task reaches an average classification accuracy of 0.95 for 10 types of behind wall motions, which is a dramatical improvement when compared to the previously reported work concerning to through wall motion classification [58], [59].

**Confusion Matrix of R-max TD**

| True \ Predicted | walk forward | walk backward | sit down | stand up | fetching | crawl forward | crawl backward | fall forward | fall backward |
|---|---|---|---|---|---|---|---|---|---|
| walk forward | 0.99 | 0.00 | 0.00 | 0.00 | 0.00 | 0.00 | 0.00 | 0.00 | 0.00 |
| walk backward | 0.00 | 1.00 | 0.00 | 0.00 | 0.00 | 0.00 | 0.00 | 0.00 | 0.00 |
| sit down | 0.00 | 0.00 | 0.90 | 0.00 | 0.07 | 0.00 | 0.01 | 0.00 | 0.01 |
| stand up | 0.00 | 0.00 | 0.00 | 0.98 | 0.02 | 0.00 | 0.00 | 0.00 | 0.00 |
| fetching | 0.00 | 0.00 | 0.07 | 0.00 | 0.92 | 0.00 | 0.00 | 0.00 | 0.01 |
| crawl forward | 0.00 | 0.00 | 0.00 | 0.00 | 0.00 | 1.00 | 0.00 | 0.00 | 0.00 |
| crawl backward | 0.00 | 0.00 | 0.01 | 0.00 | 0.00 | 0.00 | 0.99 | 0.00 | 0.00 |
| fall forward | 0.00 | 0.00 | 0.00 | 0.00 | 0.00 | 0.00 | 0.00 | 1.00 | 0.00 |
| fall backward | 0.00 | 0.00 | 0.01 | 0.00 | 0.00 | 0.00 | 0.00 | 0.00 | 0.99 |

In addition, comparative classifications with the previously reviewed RA-STFT and CRATFR feature enhancement methods are conducted. The average classification accuracy for each method is listed below in Table 2. As can be observed, the proposed R-max TFR method achieves the highest average classification accuracy with the value of 0.95. Such a performance further demonstrates that our proposed method precedes the others in motion recognition and classification performance.

Table 2

| Methods | Average Classification Accuracy |
|---|---|
| STFT from narrow band | 0.6 |
| STFT from wide band | 0.7 |
| CRATFR | 0.8 |
| R-max TF | 0.95 |

## V. Conclusion

In this paper, the motion recognition and classification task are firstly extended to through complex wall scenario. Then, a micro-Doppler feature enhancement strategy by combing high-pass filtering and R-max TFR method is proposed. Experiment verification is applied to demonstrate the effectiveness of the proposed strategy. We can conclude that the utilization of R-max TFR when analyzing the micro-Doppler effects of the behind wall motions can offer not only the applicants a micro-Doppler signature with improved quality, but also the alleviation of the processing burden. Besides, 2D-PCA based classification further proved that the proposed R-max TFR method outperforms other state-of-art micro-Doppler analysis methods.

It should be noted that the proposed method is suitable for any other possible UWB radar system, such as FMCW UWB radar, as long as the HRRP could be obtained. However, impulse-radio (IR) UWB radar is an exception since no phase information could be retrieved using the IR-UWB radar system. In addition, it does not exert any requirement for the operating frequency. Thus, it has great potentials for human-computer interaction task because the micro-Doppler signatures of hand gestures are very weak.

The method also possesses the potentials for more extreme scenarios like penetrating multiple layered walls, which might reduce the antennas numbers needed for a smart home scene.

## Acknowledgment

The author would like to give his specific thanks for Prof. Moeness Amin, Center for Advanced Communication, Villanova University, for his insight and meaningful discussions on this topic during the author's visiting stay in the center.

The author would also like to thank the China Radio Propagation Research Institute (CRPRI), Qingdao, China, for their kindness of providing us the experimental site for all the experiments conducted in this work.